\documentclass[ras]{agu2001}
\usepackage [dvips]{graphicx}
\newcommand{\dmu}{${\rm pc \ cm ^ {-3}}$}
\def\la{\mbox{\raisebox{-0.1ex}{$\scriptscriptstyle \stackrel{<}{\sim}$\,}}}

\authorrunninghead{BHAT ET AL.}

\titlerunninghead{RFI Identification and Mitigation}

\begin{document}

%
%
%
%
%

%
%

\title{RFI Identification and Mitigation Using Simultaneous Dual Station Observations}
%

%
%


\author{N. D. R. Bhat}
\affil{Massachusetts Institute of Technology, Haystack Observatory, Westford, MA~01886}

\author{J. M. Cordes}
\affil{Astronomy Department and NAIC, Cornell University, Ithaca, NY~14853}

\author{S. Chatterjee}
\affil{National Radio Astronomy Observatory, 1003 Lopezville Road, Socorro, NM 87801}

\author{T. J. W. Lazio}
\affil{Remote Sensing Division, Naval Research Laboratory, Washington, DC 20375-5351}

\begin{abstract}
RFI mitigation is a critically important issue in radio astronomy 
using existing instruments as well as in the development of 
next-generation radio telescopes, such as the Square Kilometer 
Array (SKA). Most designs for the SKA involve multiple stations 
with spacings of up to a few thousands of kilometers and thus can 
exploit the drastically different RFI environments at different 
stations.  As demonstrator observations and analysis for SKA-like 
instruments, and to develop RFI mitigation schemes that will be 
useful in the near term, we recently conducted simultaneous 
observations with Arecibo Observatory and the Green Bank Telescope 
(GBT). The observations were aimed at diagnosing RFI and using the 
mostly uncorrelated RFI between the two sites to excise RFI from 
several generic kinds of measurements such as giant pulses from 
Crab-like pulsars and weak HI emission from galaxies in bands 
heavily contaminated by RFI.  This paper presents observations, 
analysis, and RFI identification and excision procedures that are 
effective for both time series and spectroscopy applications using 
multi-station data.
\end{abstract}

%
%

%

\begin{article}

%
%

\section{Introduction} \label{s:intro}

Radio astronomers have begun to exploit the recent advent of
wide-bandwidth receivers and spectrometers in order to obtain higher
sensitivity as well as increased frequency coverage for spectroscopic
observations.  Consequently, observations 
often need to be made outside the (fairly narrow) frequency bands 
allocated to radio astronomy, forcing 
observers to explore various means of co-existence with other users of 
the spectrum, such as commercial and defense. In order to realize the
full potential of these wide-bandwidth receivers and spectrometers,
radio frequency interference (RFI) mitigation techniques will need to 
be devised and 
implemented at different stages within the signal path---at radio frequency
(RF) front-end, pre-correlation and post-correlation.


\begin{figure*}
\centerline{
\noindent\includegraphics[width=25pc,angle=270]{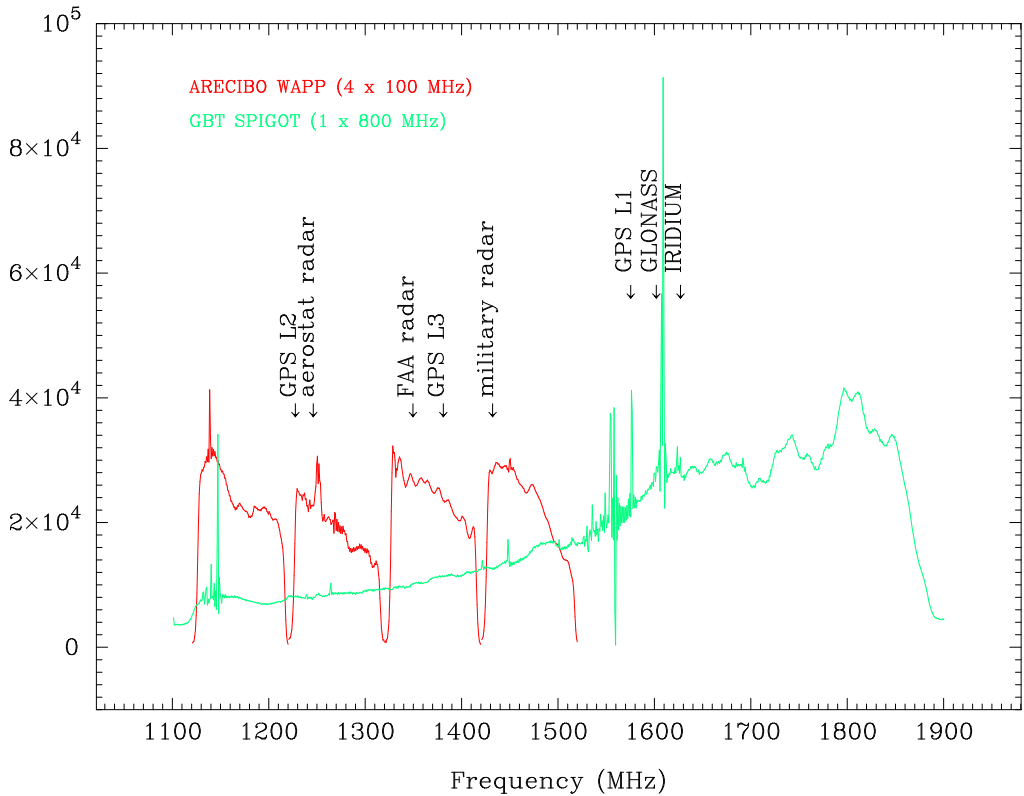}
}
\caption{Example plots showing the instantaneous RFI environments in 
the L-band frequency range at the Arecibo and Green Bank Telescopes; 
these spectrometer bandshapes are from a short integration (1 s) of 
data taken on UGC 2339. Locations of some of the most prominent RFI
sources are also marked.}
\label{f:rfi}
\end{figure*}

Simultaneously, next
generation radio telescopes are being designed and developed---such as 
the Allen Telescope Array (ATA),
the Low Frequency Array (LOFAR), the Long Wavelength Array (LWA), and
the Square Kilometer Array (SKA)---whose
designs are radically different from most existing single-dish or
interferometric telescopes. Naturally, they will need to deal with 
RFI environments that are drastically different from what observers 
have used so far. As such, RFI mitigation is recognized as 
a critically important 
issue in the development of these new class of instruments. Many of
these next-generation radio telescopes (most notably the \hbox{SKA},
the LWA and to a lesser extent LOFAR) make use of array 
stations with separations as large as hundreds to even  thousands of
kilometers; the RFI environment can be expected to 
vary significantly depending on location of the station
within the array, and the regulatory protection available 
in its vicinity. 
The same may apply to existing instruments such as the European 
VLBI Network (EVN) and the Very Long Baseline Array (VLBA), albeit
they have much fewer number of stations and are primarily used
in imaging applications.
A design incorporating a large number of array stations at large
distances from 
each other means that a possible RFI  identification and 
mitigation  technique is to exploit these large array station separations.

As demonstrator observations for SKA-like instruments, and also with 
the general goal of development of RFI excision methods that can 
potentially be applied to fast-sampled data in time and frequency, 
we recently conducted simultaneous observations with the Arecibo and 
Green Bank Telescopes. Data from such observations represent only a 
sub-class of data obtainable from SKA-like instruments, and relevant 
for non-imaging type applications envisaged with the SKA, such as 
pulsar astronomy and transient searches. As part of these observations,
data were gathered on the Crab pulsar, several beam areas on the
galaxy M33, and the galaxies UGC~2339 and UGC~2602, at the L band 
frequency range. This paper will present observations, analysis, and 
RFI identification and excision procedures that are effective for both 
time series and spectroscopy applications using the dual-station data.

\section{Arecibo-Green Bank Observations} \label{s:aogbt}

Data used for the analysis presented in this paper are taken from
observations made in November 2003 using the Arecibo and Green Bank 
Telescopes. At Arecibo,
data were recorded using the wide-band correlation spectrometer, the
Wideband Arecibo Pulsar Processor (WAPP, {\tt http://www.naic.edu/$\sim$wapp}), 
four identical units of which were used to yield
a  total bandwidth of 400 MHz that spans
the frequency range from 1120 to 1520 MHz. Data 
acquisition at the Green Bank Telescope (GBT) was done using the new 
GBT spectrometer SPIGOT card ({\tt http://www.gb.nrao.edu/GBT}), 
which is capable of a maximum bandwidth of 800 MHz, thus
covering the frequency range from 1100 to 1900 MHz. As a result
nearly 400 MHz of band is common to both the data sets.


\begin{figure*}[h]
\vskip -0.10in
\centerline{
\noindent\includegraphics[width=20pc,angle=270]{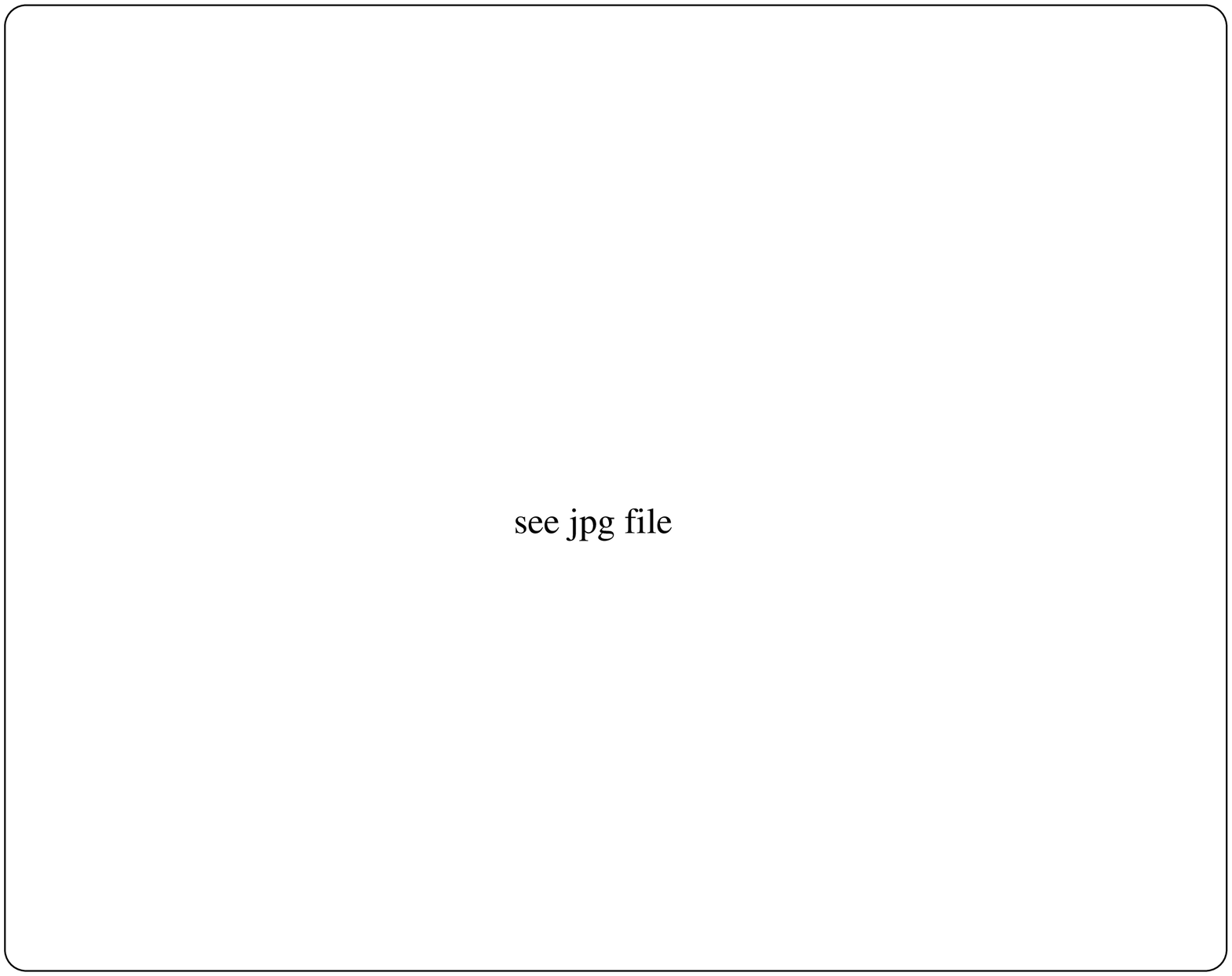}
}
\vskip -0.10in
\caption{Schematic diagram illustrating the algorithm for identification 
and excision of RFI in fast-sampled spectrometer data. Data flow shown
here follows the steps (1) to (6) as described in \S~\ref{s:alg}.}
\label{f:alg}
\end{figure*}

In order to attempt our prime goal of exploring dual-station approach
for better RFI excision techniques, specifically for spectroscopy
applications, it is essential that data recording at the two telescopes 
use identical resolutions in time and frequency. This was
feasible in our case owing to high level of flexibility and versatility 
offered by the WAPP systems, which allow the user to set arbitrary 
time and frequency resolutions. The observing parameters for each WAPP were 
chosen to exactly match that of the 800 MHz mode with the SPIGOT card. 
Data were taken at a sample interval of~82~$\mu$s, 
512 spectral channels spanning the composite  400 MHz band 
of the four WAPPs, and 1024 channels spanning the 800 MHz band of 
SPIGOT (i.e. a spectral resolution of 0.781 MHz for both data
sets).  Incidentally, by the time of these observations, Arecibo's 
Gregorian system was equipped with a new L-band wide 
receiver ({\it http://www.naic.edu/$\sim$astro/RXstatus/Lwide/Lwide.shtml})
whose design is 
identical to the one at the GBT.  Thus our observing setups at the two 
telescopes are nearly identical, in terms of both spectrometer settings 
and choice of receivers, except for the obvious difference in 
primary beam widths and achievable
sensitivities for the two telescopes.


\begin{figure*}
\vskip 0.0in
\centerline{
\noindent\includegraphics[width=25pc,angle=270]{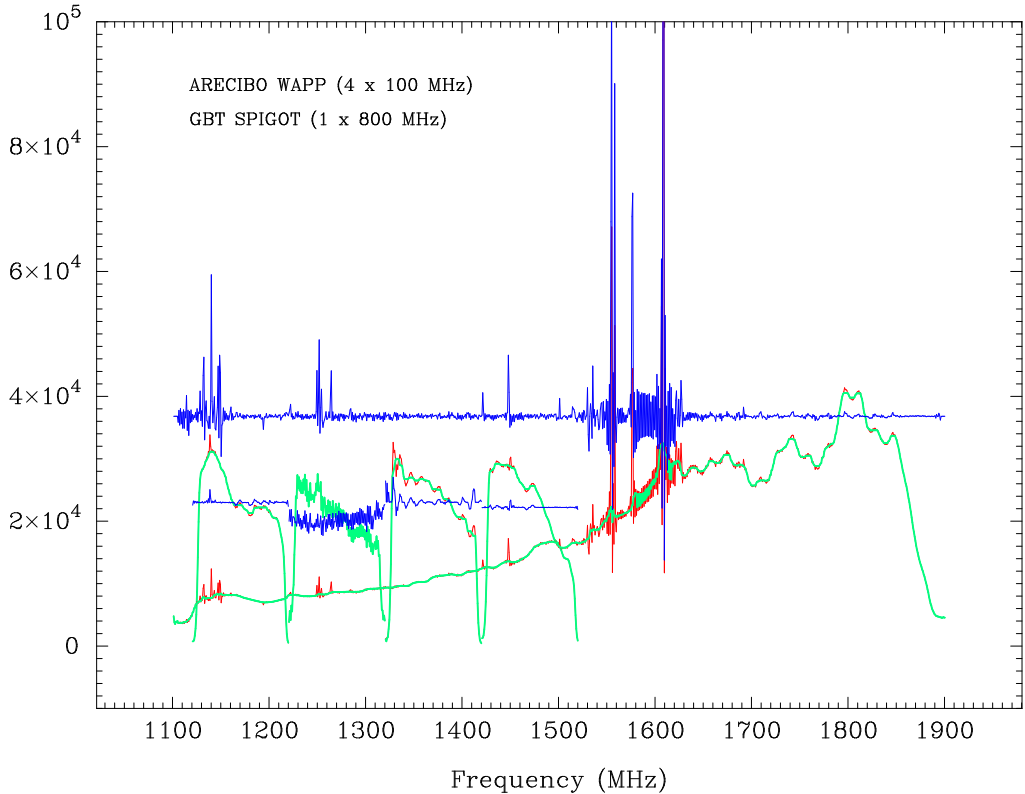}
}
\caption{Example spectrometer bandshapes from a short integration (1 s) of 
data taken on the UGC 2339 galaxy; the raw bandshapes (red solid curves) 
are overplotted with the median-filtered, smoothed versions of the bands 
(green solid curves). The curves in blue represent the ratios of the raw 
data to smoothed data, and appropriately scaled for display.}
\label{f:med}
\end{figure*}

In the remainder of the paper we describe some techniques useful
for identification and excision of RFI in (i) fast-sampled 
spectrometer data, and (ii) dual-station time series data.  
Specifically, in \S~\ref{s:alg} we focus on a generic algorithm 
that can be applied to two-dimensional data in the time-frequency 
plane (i.e.  dynamic spectrum) to generate RFI 
``masks'' for input to a data processing pipeline, and in 
\S~\ref{s:trans} we leverage the utility of multiple-subband 
and multiple-site approaches to differentiate impulsive RFI 
from real signal (events) in time series data.

\section{Algorithm for Identification and Excision of RFI 
in Fast-sampled Spectrometer Data} \label{s:alg}

Figure~\ref{f:rfi} shows examples of one-second averaged 
spectra obtained simultaneously in 
the L-band frequency range at Arecibo and Green Bank.
Examination of such data as a function of time reveals the presence 
of a variety of RFI, ranging from RFI that is narrowband 
and persistent in nature (e.g. GPS modes L1, L2 and L3) to those
that are strong and bursty (e.g. Iridium line at 1627 MHz). 
Their prominence can 
potentially lead to quite detrimental effects in 
applications such as pulsar data processing, and 
may considerably limit the sensitivity achievable 
for spectroscopy observations.

We have devised an algorithm for the identification and 
excision of RFI in fast-sampled spectrometer data. 
Such data are especially sensitively to RFIs that are 
mostly strong and impulsive in nature. The general scheme 
involves applying a two-dimensional running median filter 
to the data (e.g. Ransom et al. 2004), followed by a series 
of thresholding and excision (flagging) stages, and possibly 
in multiple passes, in order to identify regions of 
data samples in the time-frequency plane that are corrupted 
by RFI. There are several adjustable parameters at
various stages of the algorithm that can be optimized 
to yield the best results.

Note that for the sake of computational efficiency, 
as well as to ensure sufficient interference-to-noise ratio
(INR), it is important to smooth and decimate
(preferably in time) prior to application of the algorithm.
In general, the decimation factor will depend largely on the science 
application, the RFI environment at the telescope, and acceptable 
levels of data loss due to RFI excision. The algorithm essentially 
steps through the sequence as listed below (see also Fig.~\ref{f:alg}): 


\begin{figure*}[h]
\vskip -0.5in
\centerline{
\noindent\includegraphics[width=25pc,angle=270]{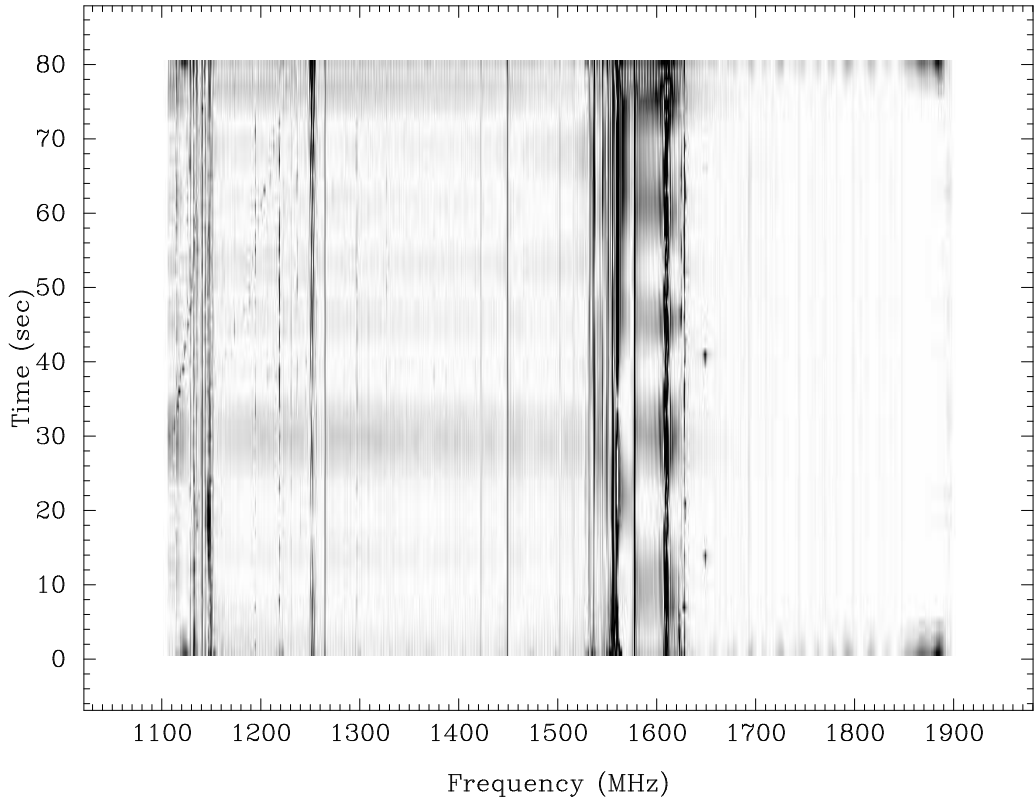}
}
\caption{Gray scale plot showing the RFI-corrupted regions 
as identified by application of a two-dimensional median filter
(5 x 5 pixels) to a short stretch of data taken on UGC 2339. 
The strength of the RFI (relative to the median bandpass level)
is shown as a logarithmic gray scale, saturating (black) at 
1.5 dB above the median power level.}
\label{f:gray}
\end{figure*}

\begin{description}

\item[(1) Data decimation]:
Start with full resolution (raw) data in the form 
of a dynamic spectrum, perform data decimation
by doing short integrations of successive
spectra in time (e.g., 0.1 to 1 s) to generate a 
time sequence of short-averaged spectra. 

\item[(2) Median filtering]:
Apply a two-dimensional median filter in time and
frequency. This is an efficient method to identify 
RFI that is impulsive in nature, either in time 
or in frequency, or both. The ratio of the dynamic 
spectra of
decimated raw data to the median-filtered data
(Fig.~\ref{f:gray}) now forms the input for further stages.

\item[(3) Thresholding in frequency]:
Identify and excise RFI that is persistent (and
narrowband) in nature. This is achieved by averaging 
longer stretches of data, followed by computation 
of its statistics (in frequency), and application of
suitable thresholding (e.g.  3-$\sigma$) to identify 
spectral channels with persistent RFI.

\item[(4) Thresholding in time]:
Identify and excise RFI that is broadband (and
transient) in nature. For this, integrate the data 
over the full band (or multiple sub-bands), followed 
by computation of its statistics (in time), and 
application of thresholding.

\item[(5) Thresholding in time and frequency]:
Identify and excise RFI that is localized in
the time-frequency plane. Perform channel-wise 
computation of the statistics of the data, and
apply thresholding in the two-dimensional plane,
to identify regions corrupted by RFI.

\item[(6) Multiple passes]:
Iterate steps (3) to (5) until a satisfactory 
level of RFI excision is achieved. 
\end{description}


\begin{figure*}[h]
\vskip -0.20in
\centerline{
\noindent\includegraphics[width=25pc,angle=270]{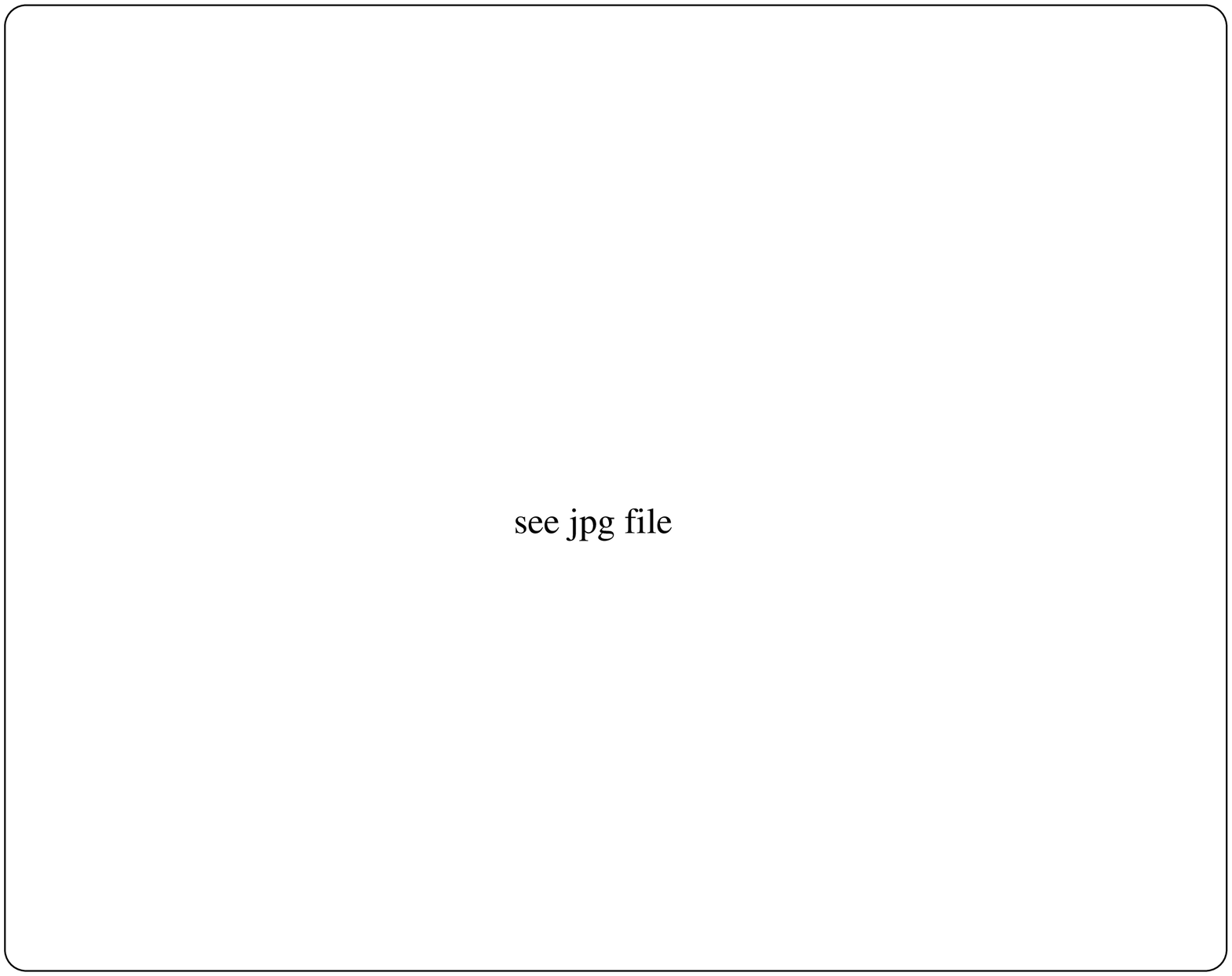}
}
\caption{Gray scale plots illustrating the application of RFI identification 
and excision algorithm on spectroscopy data; the regions in red correspond 
to the data that are identified to be corrupted with RFI. Darker regions of 
the gray scale plot indicate the residual, unexcised RFI. The strength of the
RFI is shown on a logarithmic scale, with the gray scale saturating (black)
at 1.5 dB above the median power level.}
\label{f:red}
\end{figure*}


\begin{figure*}
\vskip 0.0in
\centerline{
\noindent\includegraphics[width=25pc,angle=270]{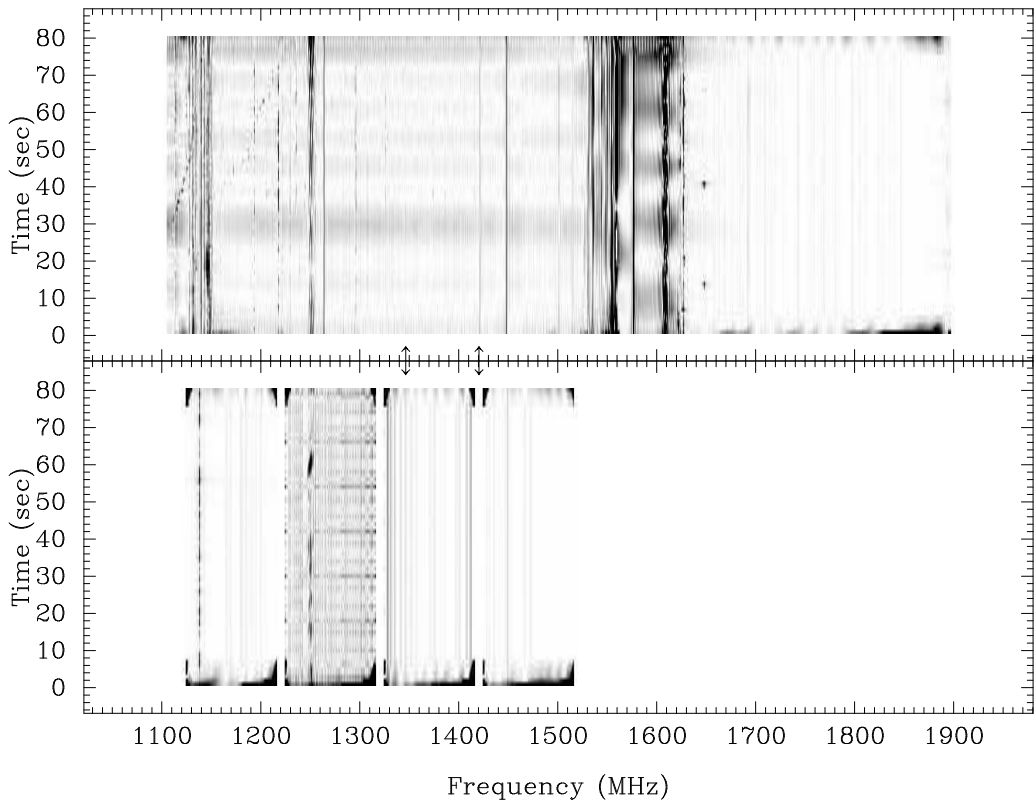}
}
\caption{Gray scale plots showing the RFI-corrupted regions of the data 
taken on UGC 2339. The strength of the RFI is shown on a logarithmic gray 
scale, saturating (black) at 1.5 dB above the median power level. 
Different types of RFI can be seen, ranging from persistent RFI confined 
to a small frequency range to those that are bursty and strong (e.g. Iridium 
line). The top panel shows the GBT-SPIGOT data, while Arecibo-WAPP data are 
shown in the bottom panel. Locations of the unseen HI emission from UGC 2339
(at 1346.5 MHz) and that of the Galactic HI emission (at 1420.4 MHz) are 
indicated by two-sided arrows in the figure.} 
\label{f:dual}
\end{figure*}

Figures \ref{f:med}--\ref{f:dual} show example plots that illustrate this
method, using the UGC 2339 data taken with the GBT-SPIGOT system. Based 
on these and the above description, it is evident that the algorithm is
primarily sensitive to impulsive type RFIs, either in time or frequency,
such as narrowband bursts or persistent RFIs confined to a few spectral
channels. The underlying assumption is that most real celestial signals are
characteristically different, and may even show signatures of dispersion
and/or scattering (e.g. frequency structure from diffractive
interstellar scintillation). While our method is primarily intended as an RFI 
finding algorithm (and not a transient event detection scheme), 
certain fast transients of broadband nature
may be classified as RFI by this scheme. An ideal way to circumvent
this may be to capture such ``localized RFI events'' and examine them
in detail later on using appropriate matched-filter-like schemes for 
possible real signals. A detailed implementation of such an algorithm 
is beyond the scope of this paper. 

The above scheme offers the choice of several control parameters 
that can be optimized to yield the best results. These include 
the window for median filtering, thresholds adopted at various 
stages of the algorithm and the number of iterations performed.  
The size of the median filter window is largely dictated by the 
resolution of data (in time and frequency), and the types 
of RFI that are present in the data. Our trials suggest 
use of a 5x5 window (i.e. the pixel under consideration is 
replaced by the median of all 25 elements in the 5x5 square 
neighborhood box in the time-frequency plane, including the 
diagonal elements) as an optimal choice to obtain the best 
results for the data taken with the SPIGOT and WAPP spectrometers. 
We also
emphasize that our use of a median filter is merely to identify
RFI-corrupted intensity samples in the data, and hence it does 
not alter any statistical characteristics of the original data. 
The choice of thresholds can be based either on the statistics
of the data samples, or in terms of a percentage increase of 
the mean bandpass level. Finally, the number of iterations
influences the quality of RFI excision; for example, the use 
of fewer iterations may potentially result in some residual 
unexcised RFI in the data, the extent of which may serve as 
a useful figure of merit to assess the quality of RFI excision 
for a given set of parameters. As an end product, the algorithm 
effectively generates two-dimensional RFI ``masks'' which
can be inputted to the various data processing pipelines.

In general, the choice of optimal thresholds and number of 
iterations will depend on the degree of RFI contamination.
For the data sets shown in Figs.~\ref{f:med}--\ref{f:dual}, 
we find typically 3 to 4 thresholding passes (i.e. 1-2 to
identify and excise spectral channels of persistent RFI,
and the remainder 1-2 for the RFI-corrupted sections of
data in the time-frequency plane) to be sufficient for 
eliminating a large fraction of the RFI.  Generally, in
our case most narrowband RFI are eliminated by using a 
threshold of 0.2 to 0.6 dB (i.e. 5 to 15\%) above the 
median power level, while a threshold of 3--4 $\sigma$ 
seems to work well for RFI-excision in the two-dimensional 
time-frequency plane. We however emphasize that the choice 
of thresholds and number of passes are, in general, largely 
dictated by the spectrometer and receiver combination. 
Thus the optimal scheme for RFI excision may differ for 
the data taken with a different receiver or telescope 
and/or using a different spectrometer backend.

\subsection{Extension to Dual-station Data} \label{s:ext}

As stated in the previous sections, one of the original goals 
of these demonstrator observations was to exploit the dual-site 
nature of the data for its effectiveness to do better spectroscopy, 
especially in regions of the band heavily contaminated by RFI. 
We applied the above-described RFI-excision algorithm on the UGC
data set from each telescope separately, in order to examine the 
difference in the RFI environments at the two telescopes. 
Figure~\ref{f:dual} shows the degree of RFI 
contamination at the two telescopes as a function of time 
and frequency for the data sets of UGC 2339. As seen from 
the figure, the RFI environment is relatively cleaner at 
Arecibo in the lower 100 MHz of the band, while it is 
considerably worse in the adjacent 1220--1320~MHz range. 
In general, RFI is seen to be largely uncorrelated between 
the two telescopes, and in principle, such uncorrelated RFI 
can easily be identified and excised using dual-site RFI 
mask data generated by the algorithm. For this galaxy, the 
HI emission is expected near 1346.5 MHz with a line width 
of approximately 2.1 MHz. Its apparent lack of detection 
in our data can be attributed to insufficient data length 
and coarse spectral resolution (0.781 MHz). 
However the data reveal the presence of 21 cm emission 
(at 1420.4 MHz) due to our own Galaxy.


\begin{figure*}[b]
\vskip 0.20in
\centerline{
\noindent\includegraphics[width=18pc,angle=270]{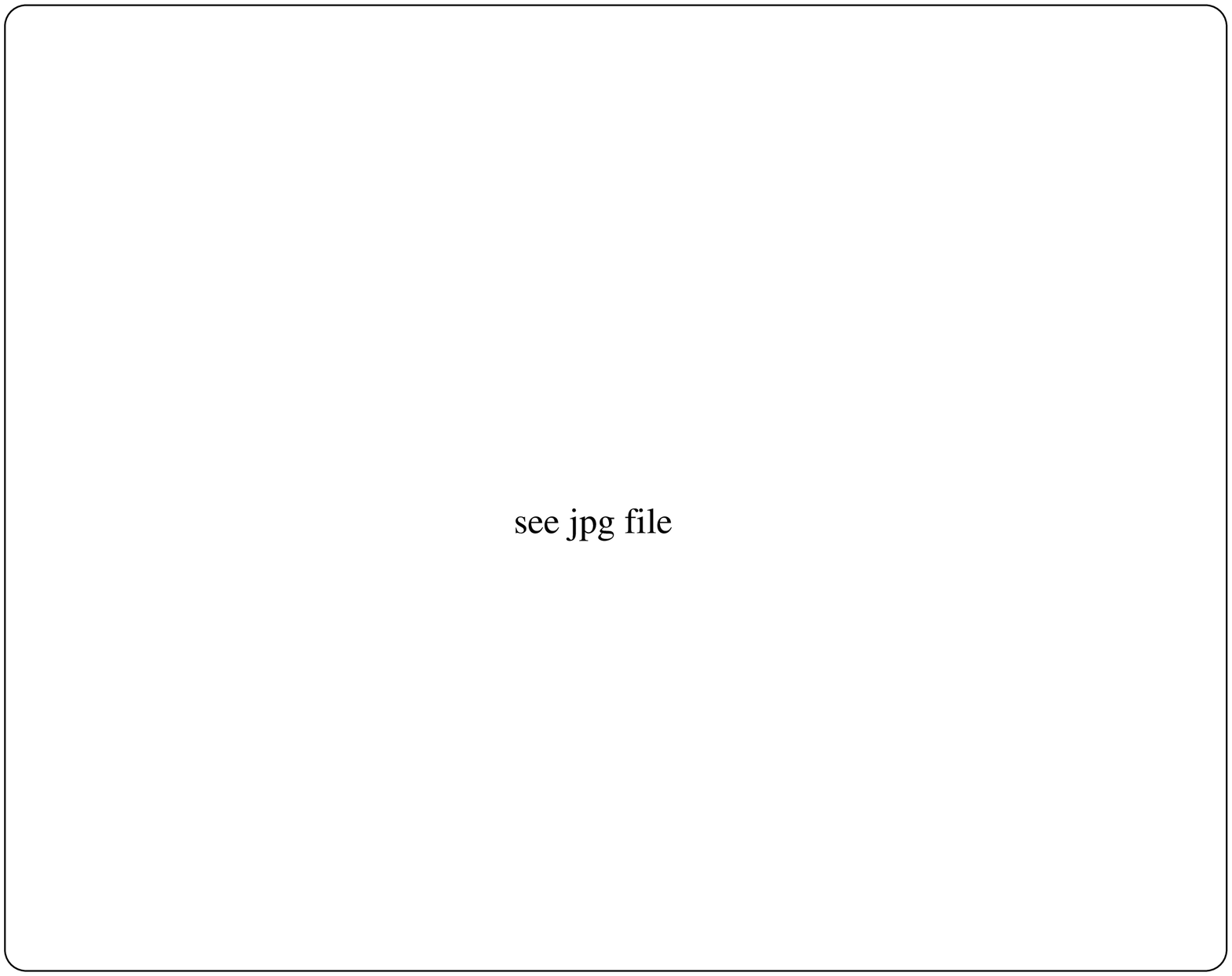}
}
\vskip 0.25in
\caption{Schematic illustration of the transient event (e.g. giant pulse) 
detection method described in \S~\ref{s:gpdet}. Data at the full resolution 
(in time and frequency) are converted into dedispersed time series, and 
searched for possible ``events.'' The event lists from multiple sub-bands 
or multiple sites can be compared to differentiate real events from spurious 
ones due to RFI.} 
\label{f:event}
\end{figure*}

\section{Transient Detection in Time Series Data: RFI vs Real Events}
\label{s:trans}

Detection of transient signals is an important science goal for several
upcoming and planned large telescopes such as the \hbox{ATA},
\hbox{LOFAR}, the Mileura Wide-field Array (MWA), the \hbox{LWA}, 
and the \hbox{SKA}, all of which are planned to have
relatively wide fields of view, a key
requirement for transient search.  The Crab pulsar emits ``giant
pulses,'' radio bursts whose amplitude can exceed the average pulse
amplitude by orders of magnitude (e.g., Lundgren et al.~1995, and
references within).  There have been a variety of recent 
attempts to search for giant pulses from both Galactic 
and extragalactic objects (McLaughlin \& Cordes 2003; Johnston \& Romani
2003). One major difficulty with these searches is the prevalence of 
impulsive RFI, which often mimics giant-pulse-like signals in the 
time series data. Strong RFI can also potentially mask even the 
characteristically dispersed pulses from pulsars. Most new generation 
telescopes will offer unique means of circumventing this problem, via 
use of multiple sites and/or simultaneous multiple beams.

Our demonstrator observations described in \S~\ref{s:aogbt} are well suited
for the exploration and development of techniques that could 
help discriminate between real signals and impulsive RFI.  Specifically, 
our data taken on the Crab pulsar provide an excellent test bed for this 
purpose. We describe the single-pulse (giant-pulse) search method 
(Cordes \& McLaughlin 2003) as an example of transient event detection 
schemes applicable to fast-sampled data (see Fig.~\ref{f:event}), and 
demonstrate the leverage 
of dual-site and multiple-subband approaches for discriminating 
``real events'' from those that are caused by impulsive RFI.

\subsection{Transient Event (Giant Pulse) Detection in Fast-sampled Data}
\label{s:gpdet}

The single-pulse (giant-pulse) search method used for our analysis 
(Fig.~\ref{f:event}) is
essentially the same as that described in Cordes \& McLaughlin (2003)
and used by Cordes et al. (2004).  The 
raw data are first converted into a filterbank stream, retaining the original 
resolution in time and frequency. These are then dedispersed by summing 
over frequency channels while taking into account time delays associated
with plasma dispersion in the interstellar medium (ISM). The dedispersion
is performed at one or many dispersion measures (DMs) depending on whether the
object is of a known DM (e.g. the Crab pulsar) or unknown DM. 
Often knowledge of the DM range expected toward an object (e.g. 60--80 \dmu
toward the M33 galaxy; Cordes \& Lazio 2002) may help to narrow down 
the search in the DM parameter space. 
For reference, similar analysis is also performed for the case DM=0. 

The dedispersed time series 
data are first analyzed with the original time resolution, and then
progressively smoothed and decimated by factors of two in order to 
approximately match filter to pulses with different widths. 
Individual pulses (i.e. events) and their occurrence times are then 
identified by selecting intensity samples that exceed the mean 
level by a specified threshold (e.g., 3 to 5$\sigma$). This process 
can (optionally) be performed iteratively by excluding the detected 
events (or the corresponding intensity samples) from previous 
iterations, until no more events are detected. These identified 
events can then be examined for detailed characteristics such as 
their strength, spatial grouping, time-frequency signature, etc. 

While the above method appears to be effective to search for 
giant-pulse-like events from objects with known DMs, 
identification of real events becomes more difficult owing to
false positives for the more 
general case of an unknown DM and an unknown type of signal, 
for which many more statistical trials must be done.
In the remainder of the section we illustrate the 
leverage of some simple-minded, yet powerful methods of 
discriminating real events from spurious ones that may be
generated by RFI.

\subsubsection{Dual Station Technique}
\label{s:dual}

This is the simplest possible illustration of a more general multiple-site
based approach that can potentially be used to differentiate real events 
from spurious ones caused by RFI. Assuming that much of the RFI is largely 
local to the site, and consequently likely to be uncorrelated between 
the two sites, a simple technique such as examining for coincidence vs 
anti-coincidence of the detected events can be used to filter out 
the events that are more likely to be real. In order to apply this 
method, it is essential that data from the two telescopes be
processed in an identical manner, and that identical criteria 
are applied for the generation of event lists. The times of 
occurrence of the events are corrected for (a) any dispersion 
delays due to differences in the reference frequencies used for 
the dedispersion process; (b) any instrumental delays; and (c) 
differences in the arrival times of the signal at the telescopes, 
before attempting the comparison. The events are then examined for 
the simultaneity as well as similarity in terms of properties, and 
those that are unambiguously detected by both the telescopes are 
categorized as likely real events. 

\begin{figure*}[b]
\vskip 0.0in
\centerline{
\noindent\includegraphics[width=18pc,angle=270]{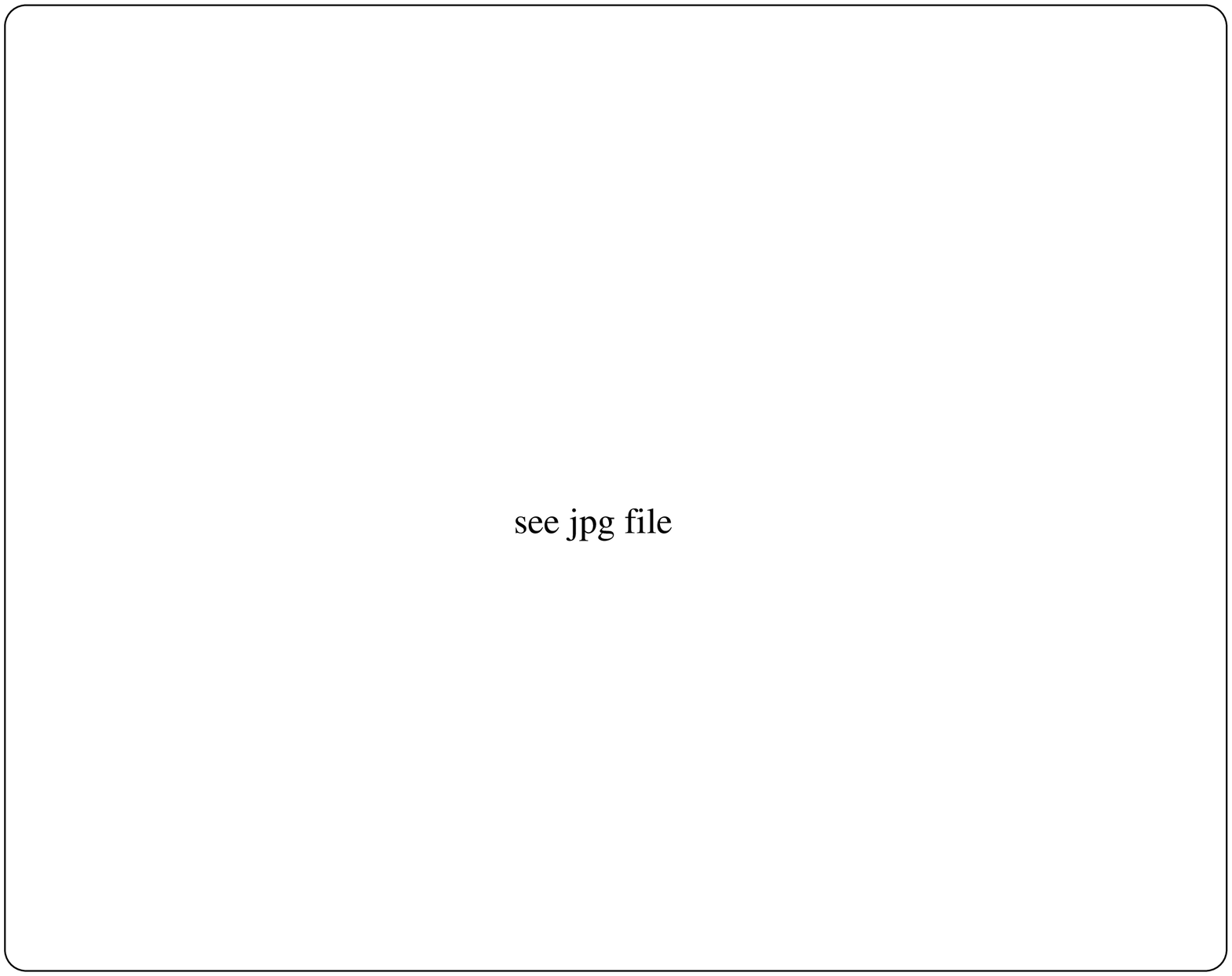}
}
\vskip 0.0in
\caption{Dedispersed time series of the Crab pulsar from the Arecibo-Green
Bank observations at L-band; data shown are for a short duration of 8 s.
The top panel is the Arecibo data and the bottom one is the GBT data. 
Locations of the detected events (in Green Bank data) are indicated by 
the upward arrows; the events that are common to both the data sets 
correspond to positive identifications of giant pulses, and those that 
are unlikely to be real are marked by the `?' symbol.}
\label{f:crab}
\end{figure*}

Despite its simplicity, there are some caveats for this method. 
As the signal-to-noise ratio (S/N) of a detected event depends on 
the sensitivity of the instrument, it is quite possible many real 
events with only marginal detections at the more sensitive telescope
may be classified as spurious ones. Such ambiguity can be minimized
by appropriate scaling of the detection thresholds for uniformity of
the detection criteria in terms of, for example, the minimum detectable 
flux density. Figure~\ref{f:crab} shows an example segment of dual-station 
time series data of the Crab pulsar; both data sets
show several large-amplitude spikes. Using the simplest criterion
that ``events'' must appear in both data sets to be considered 
real (i.e. giant pulses), we identify a few spurious events in 
the GBT data that are likely caused by impulsive RFI. Further, the 
pulse-amplitude ratios appear to be slightly different for the two 
data sets, and this is most likely due to the difference in spectrometer 
bandwidths (400 and 800 MHz respectively for the Arecibo and 
Green Bank data). For data sets of identical bandwidths, the 
pulse-amplitude ratios match within a few percent level.


\subsubsection{Multiple Subband Technique} \label{s:sub}

This method is especially suitable for observations that employ large
spectrometer bandwidths (Arecibo's WAPP and the GBT's SPIGOT are
good examples). In such cases, the RFI environment may vary considerably 
across the frequency range of observation, ranging from relatively clean 
parts of the band with little or minimal RFI to those that are marked by
a highly RFI-rich environment. Often strong and bursty types of RFIs can 
potentially mask even the brightest giant pulses from Crab-like pulsars. 
Although this can be addressed in principle (or at least minimized) by 
the application of suitable RFI identification and excision algorithms 
(such as the one presented in \S~\ref{s:alg}) prior to processing, 
and the integration of the generated RFI ``masks'' into the data reduction 
pipeline, such schemes may also be computationally expensive, especially 
when dedispersion needs to be performed over many trial DMs. With some 
trade-off in the achievable sensitivity for the event detetcion, an 
alternative scheme that makes use of multiple sub-bands (preferably
equally-sized) can be applied to such data. The data are split 
into several contiguous sub-bands, each of which is processed 
individually for application of the event detection algorithm. 
The multiple event lists generated in this manner are then examined
for simultaneity (i.e. coincidence vs anti-coincidence approach 
between sub-bands), as well as similarity of the properties, 
after taking into account the time delay due to differences 
in the reference frequencies used for dedispersion.  
The events that have correspondence in two or more 
sub-bands are then classified as likely real events. 

\begin{figure*}[b]
\vskip 0.0in
\centerline{
\noindent\includegraphics[width=20pc,angle=270]{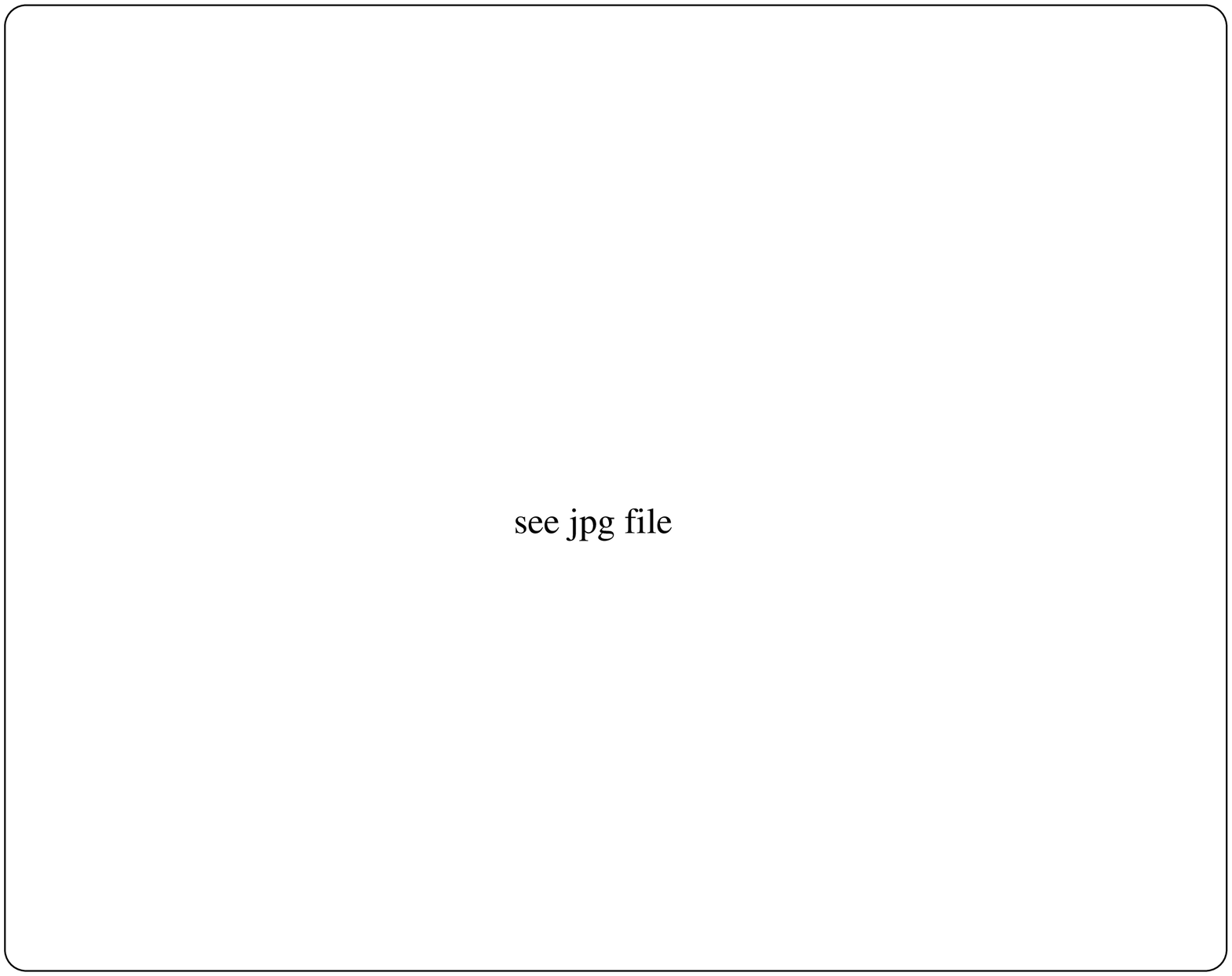}
}
\caption{Time series of ``events'' (i.e. plots of S/N vs sample number) 
detected by the single-pulse detection algorithm as described in 
\S~\ref{s:gpdet}; top to bottom panels on the left correspond to 
100 MHz bandwidth chunks spanned by four units of the WAPP, i.e. 
WAPP1 (1120--1220 MHz), WAPP2 (1220--1320 MHz), WAPP3 (1320--1420 MHz) 
and WAPP4 (1420--1520 MHz). Right panel: coincident events (green) 
for the different sub-band pairs WAPP1-WAPP2, WAPP1-WAPP3 and 
WAPP1-WAPP4 (top to bottom), overplotted on the time series 
of events detected in the WAPP1 sub-band (left, top panel).}
\label{f:cross}
\end{figure*}

We illustrate this scheme using example data sets taken on the Crab pulsar 
in the frequency range  from 1120 to 1520 MHz with the Arecibo
WAPP system. In this case, four identical units of the multi-WAPP system, 
each capable of a maximum of 100 MHz bandwidth---spanning a total of 
400 MHz band---make a logical split of the data into four contiguous
sub-bands. Data from the four sub-bands are processed separately
using a uniform set of event detection criteria in order to generate 
the respective event lists, which are shown in Fig.~\ref{f:cross}. 
Clearly, the number of detected events (for an assumed 10$\sigma$ 
threshold) is a strong function of the level of RFI contamination 
within the band, resulting in nearly two orders of magnitude larger 
number of events in the WAPP unit that spans the 1220 to 1320 MHz 
range (see also Figs.~\ref{f:rfi}~and~\ref{f:med}). 
Examination of these event lists for simultaneity or correspondence
between different sub-band pairs yields a much fewer number of events 
that are common. In fact the least number of events is detected for the 
combination that includes the RFI-rich band (1220--1320 MHz) of WAPP2. 
A plausible reason for such a disparity may be an apparent increase 
in the noise rms resulting from the high degree of RFI contamination, 
leading to many real events escaping through the event detection algorithm.


Despite its simplicity and demonstrated success, some consequences and 
caveats are in order for this method. First, this apparent efficiency 
comes at the cost of somewhat larger processing requirements; i.e. 
N data streams to deal with at the post-dedispersion stage (where N 
is number of sub-bands), and in the case of many-DMs, this implies 
roughly N-fold increase in processing requirements. Further, some 
caution is needed in the interpretation of uncorrelated or partially 
correlated events between different sub-bands. For example, diffractive 
interstellar scintillations 
modulate the frequency structure of intensity at a given 
time, and this may potentially lead to apparent lack of simultaneity
of real events. Indeed some knowledge of the expected scintillation 
bandwidth (predictable for a given DM and line of sight with some 
reasonable accuracy) can be integrated into event detection criteria. 
For instance, the Crab pulsar is known to show scintillation structures 
$\la$1 MHz bandwidth at L-band frequencies (Cordes et al. 2004). 
Moreover
the scintillation characteristics vary with the epoch of observation. 
This can potentially be addressed by use of more complex and stringent
criteria for filtering out real events, such as a simultaneity check 
of events in more than two sub-bands, and possibly between multiple
sites. In any case, this method illustrates a feasible and efficient
scheme that can be applied to data from wide-bandwidth observations.

\section{Summary and Conclusions} \label{s:conc}

We have developed and tested an algorithm for identification and excision of 
RFI in fast-sampled spectrometer data. The basic principle involves the 
application of a two-dimensional median filter, followed by a series of 
thresholding and excision performed in stages, and possibly in multiple 
passes, to yield the best results. The method is sensitive to a wide variety 
of RFI and can be used to generate RFI masks for input into 
applications such as pulsar data processing or spectroscopy in bands 
heavily contaminated by RFI.

The prevalence of impulsive RFI has long been recognized 
as a major difficulty 
in transient event detection schemes applicable to time series data. Using 
data from our recent Arecibo-Green Bank simultaneous observations (L band) 
of the Crab pulsar, we have explored effectiveness of multi-site and 
multi-subband approaches to help discriminate between 
real events and the spurious ones due to RFI. The basic idea
involves examining the coincidence vs anti-coincidence of detected events 
after taking into account the dispersion, instrumental and other delays.
The subband technique appears to be efficient for filtering out real 
signals from the likely forest of events in the sub-bands with RFI-rich 
RFI environments. In general the combination 
of multiple-site and multiple-subband offers great promise in this arena.
Applications to specific science goals, such as detections and searching 
for giant pulses and other fast transients, are deferred to a subsequent 
paper. This will make use of our existing dual-site data on Crab and M33, 
as well as data from future such observations and the upcoming multibeam 
pulsar surveys at Arecibo.

%
%

\begin{acknowledgments}
We are grateful to Karen O'Neil and David Kaplan for all their guidance 
and assistance in the planning and conducting of these observations, and to
Scott Ransom for the PRESTO software package which was used for the initial 
analysis of these data.  This work was supported by NSF grant AST
0138263  to Cornell University for technology development for the SKA. 
The Arecibo Observatory is operated by Cornell
University under a cooperative agreement with the NSF.  
The National Radio Astronomy Observatory is a
facility of the NSF operated under cooperative
agreement by Associated Universities, Inc.  
NDRB is supported by an MIT-CfA Fellowship at Haystack Observatory.
Basic research in radio astronomy at the NRL is supported by 
the Office of Naval Research.
\end{acknowledgments}

\end{article}
\end{document}